 \documentclass[prl,aps,twocolumn,showpacs]{revtex4}

\usepackage[dvips]{graphicx}
\usepackage{dcolumn}

\newcommand{\be}{\begin{equation}}
\newcommand{\ee}{\end{equation}}
\newcommand{\bea}{\begin{eqnarray}}
\newcommand{\eea}{\end{eqnarray}}
\newcommand{\nn}{ \nonumber}
\newcommand{\ds}{\displaystyle}
\begin{document}

\title{ Fermi-liquid effects in the transresistivity in quantum Hall double layers near $\nu= 1/2 $}

\author{Natalya A. Zimbovskaya }

\affiliation{Department of Physics and Astronomy, St. Cloud State 
University, 720 Fourth Avenue South, St. Cloud, MN 56301, USA;\\
Urals State Mining University, Kuibysheva Str. 30, Yekaterinburg, Russia, 620000  }

\date{\today}

 \begin{abstract}
 Here, we present theoretical studies of the temperature and magnetic field dependences of the Coulomb drag transresistivity between two parallel layers of two dimensional electron gases in quantum Hall regime near half filling of the lowest Landau level. It is shown that Fermi-liquid interactions between the relevant quasiparticles could give a significant effect on the transresistivity, providing its independence of the interlayer spacing for spacings taking on values reported in the experiments. Obtained results agree with the experimental evidence. 
  \end{abstract}
\vspace{1mm} 


\maketitle


During the last decade double-layer two-dimensional electron gas (2DEG)
systems were of significant interest due to many remarkable phenomena they
exhibit, including so called Coulomb drag. In Coulomb drag experiments two
2DEGs are arranged close to each other, so that they can interact via
Coulomb forces. A current $ I $ is applied to one layer of the system, and
the voltage $ V_D $ in the other nearby layer is measured, with no current
allowed to flow in that layer. The ratio $- V_D/I $ gives a
transresistivity $ \rho_D $ which characterizes the strength of the
effect. The physical interpretation of the Coulomb drag is that momentum
is tranferred from the current carrying layer to the nearby one due to
interlayer interactions \cite{one,two,three}.

It was shown theoretically \cite{four,five} and confirmed with experiments
\cite{five} that the transresistivity between two 2DEGs in quantum Hall
regime at one half filling of the lowest Landau level for both layers is
proportional to $ T^{4/3} \ (T $ is the temperature of the system) which
is quite different from the temperature dependence of $ \rho_D $ in the
absence of the external magnetic field applied to 2DEGs. This
temperature dependence of the drag at $ \nu
= 1/2 $ originates from the ballistic contribution to the
transresistivity. The latter reflects the response of the two-layer system
to the driving disturbance of finite wave vector $ \bf q $ and finite
frequency $\omega $ when considering scales are smaller than the mean free
path $ l $ of electrons $ (ql \gg 1) $, and times are shorter than their
scattering time $ \tau \ (\omega \tau \gg 1)$ \cite{six}.

In further experiments \cite{seven} the Coulomb drag was measured between
2DEGs where the layer filling factor was varied around $ \nu = 1/2. $ The
transresistivity was reported to be enhanced quadratically with $ \Delta\nu = \nu - 1/2. $ It was also reported that the curvature of the enhancement depended on temperature but it was insensitive to both sign of $ \Delta \nu $ and distance $ d $ between the layers. The present work is motivated with these experiments
of \cite{seven}. We calculate the transresistivity between two layers of 2DEGs subject to a strong magnetic field which provides $ \nu $ close  to $ 1/2 $ for both layers. 

We start from the well-known expression \cite{one,three} which relates the Coulomb drag transresistivity to density-density components of the polarization in the layers $ \Pi_{(1)} (\bf q,\omega) $ and $ \Pi_{(2)} (\bf q,\omega) :$ 
    \bea 
 \rho_D &=& \frac{1}{2(2\pi)^2} \frac{h}{e^2} \frac{1}{Tn^2} \int
\frac{q^2 d \bf q}{(2\pi)^2} \int \frac{\hbar d \omega}{\sinh^2 (\hbar\omega/2T)}
      \nn \\ \nn\\ &&
\times \big|U ({\bf q}, \omega)\big|^2 {\mbox {Im}} \Pi_{(1)}({\bf
q,}\omega) {\mbox {Im}} \Pi_{(2)}({\bf q,}\omega).
                 \eea
 Here, $ U ({\bf q}, \omega)$ is the screened interlayer Coulomb
interaction, and electron densities in the layers are supposed to be equal$ (n_1 = n_2 = n) .$

 Within the usual Composite Fermion (CF) approach \cite{eight} a single layer polarizability describes that part of the density-current electromagnetic response which is irreducible with respect to the Coulomb interaction. Adopting for simplicity the RPA, we obtain the following expression for the $ 2 \times 2 $ polarizability matrix:
  \be 
\Pi^{-1} = (K^0)^{-1} + C^{-1} .
 \ee
 Here, the matrix $ K^0 $ gives the response of noninteracting CFs and C is the Chern-Simons interaction matrix. Assuming for certainty the wave vector $ \bf q $ to lie in the $ "x"$ direction we have:
 \be 
\mbox{ C\/} \mbox{= }
\left(
\begin{array}{cc}
\mbox{\/0} & \mbox{\/\ $\ds\frac{iq}{4\pi\hbar}$}\nonumber\\
\mbox{\/ $-\ds\frac{iq}{4\pi\hbar}$} &  \mbox{\/0}
\end{array} \right)\mbox{.} 
  \ee
 Starting from the expression (2) we arrive at the following results for the density-density response function $ \Pi_{00(i)} ({\bf q}, \omega ):$
   \bea 
 && \Pi_{00(i)} ({\bf q}, \omega ) =
 \Pi_{(i)} ({\bf q}, \omega)
          \nn \\ \nn\\&
 =& \frac{K_{00 (i)}^0 ({\bf
q,}\omega)}{\displaystyle 1 -
\frac{8i\pi\hbar}{q} 
K_{01(i)}^0 ({\bf q,}\omega) -
\bigg(\frac{4\pi\hbar}{q} \bigg)^2 
\Delta_{(i)}  ({\bf q,}\omega)}
 .                \eea
  Here
 \be 
 \Delta_{(i)}({\bf q,}\omega) 
= K_{00(i)}^0 ({\bf q,}\omega) K_{11(i)}^0
({\bf q,}\omega) + \big(K_{01(i)}^0 ({\bf q,}\omega)\big)^2.
                       \ee
 Within the RPA response functions included in Eqs. (4), (5) are simply
related to components of the CF conductivity tensor $ \tilde \sigma $ \cite{eight}:
  \bea 
 \frac{1}{\tilde \sigma_{xx}^{(i)} ({\bf q,}\omega)} &=& \frac{iq^2}{\omega
e^2} \left [\frac{1}{K_{00(i)}^0 ({\bf q,}\omega)} -\frac{1}{K_{00(i)}^0
({\bf q,}0)} \right]; 
    \nn \\ \nn\\
{\tilde \sigma_{yy}^{(i)}({\bf q,}\omega)} &=& -\frac{i e^2}{\omega} 
\bigg [{K_{11(i)}^0 ({\bf q,}\omega)} - {K_{11(i)}^0 ({\bf
q,}0)} \bigg ]; 
   \nn \\\nn\\
{\tilde \sigma_{xy}^{(i)} } &=& - {\tilde \sigma_{yx}^{(i)} } =
\frac{ie^2}{q} {K_{01(i)}^0 ({\bf q,}\omega)}.
                 \eea

 To proceed we calculate the components of the CF conductivity at $ \nu $
slightly away from $ 1/2 .$ In this case CFs experience a nonzero
effective magnetic field $ B_{eff} = B - B_{1/2}.$ We concentrate on the
ballistic contribution to the transresistivity, so we need asymptotics for
the relevant conductivity components applicable in a nonlocal $ (ql \gg 1)
$ and high frequency $ (\omega \tau \gg 1) $ regime. Corresponding
expressions for $ \tilde \sigma_{ij} $ were obtained in earlier works
\cite{eight}. However, these results are not appropriate for our analysis
for they do not provide a smooth passage to the $ B_{eff} \to 0 $ limit at
finite $ \bf q.$ Due to this reason we do not use them
in further calculations. To get a suitable
approximation for the CF conductivity we start from the standard solution
of the Boltzmann transport equation for the CF distribution function. This
gives us the following results for the CF conductivity components for a
single layer \cite{nine}:
  \bea 
   \tilde \sigma_{\alpha \beta} &=& \frac{m^* e^2}{(2\pi \hbar)^2}
\frac{1}{\Omega} \int_0^{2\pi}  d \psi v_\alpha (\psi) \exp \left [\! -\frac{iq}{\Omega} \int_0^\psi  v_x(\psi'')d \psi'' \right] 
  \nn\\\nn\\ &&
\times \int_{-\infty}^\psi  v_\beta (\psi') \exp \left
[\frac{iq}{\Omega} \int_0^{\psi'}  v_x (\psi'') d \psi'' \right.               
    \nn\\\nn\\ &&
 \left. +\frac{1}{\Omega \tau} (\psi' - \psi)(1 - i \omega \tau)\right ] d\psi'.
   \eea
 Here, $  m^*, \ \Omega $ are the CF effective mass and the cyclotron
frequency at the effective magnetic field $ B_{eff}; \ \psi $ is the
angular coordinate of the CF cyclotron orbit. Now we carry out some formal
transformations of this expression (7) following the way proposed
before \cite{nine,ten}. First, we expand the CF velocity components $ v_\beta (\psi') $ in Fourier series:
  \be 
v_\beta (\psi') = \sum_k v_{k\beta} \exp (i k \psi').
                 \ee
 Substituting this expansion (8) into (7) we obtain:
  \bea 
  \tilde \sigma &=& \frac{m^* e^2}{(2 \pi \hbar)^2} \sum_k v_{k\beta}\int_0^{2\pi} d \psi v_\alpha (\psi) \exp (ik\psi)
   \nn\\\nn\\ &&
 \times \int_{-\infty}^0 \exp \bigg [\big(ik\Omega - i\omega +
\frac{1}{\tau} + iqv_x (\psi)\big) \theta
  \nn\\\nn\\ &&
+ iq \int_0^\theta \big(v_x (\psi + \Omega \theta') - v_x
(\psi)\big) d\theta' \bigg ] d \theta
  \eea
 where $ \theta = (\psi' - \psi)/\Omega. $

Then we introduce a new variable $ \eta $ which is related to the 
variable $ \theta $ as follows:
 \bea 
    \eta & =& \big(ik \Omega - i\omega + \frac{1}{\tau} + iqv_x
(\psi) \big) \theta
      \nn\\&&
+ iq \int_0^\theta \big [v_x (\psi + \Omega \theta') - v_x (\psi) \big ] d \theta' ,
                 \eea
 and we arrive at the result:
  \bea 
    \tilde \sigma_{\alpha\beta} &=& \frac{i m^* e^2}{(2\pi\hbar)^2} \sum_k v_{k\beta} \int_{-\infty}^0 e^\eta d \eta
        \nn\\\nn\\ \!\! &&\!
\times \!\int_0^{2\pi} \!\!\frac{v_\alpha (\psi) \exp (ik\psi)}{\omega + i/\tau - k \Omega - qv_x (\psi + \Omega \theta)} d \psi
    .             \eea
 Under the conditions of interest $\omega \tau \gg 1, \ ql \gg 1,$ and also assuming that the filling factor is close to $ \nu = 1/2,$ so that $ qv_F \gg \Omega \ (v_F $ is the CFs Fermi velocity), the variable $ \theta $ is
approximately equal to $ \eta \tau (1 + iql \cos \psi + i k\Omega \tau - i \omega \tau)^{-1}.$ Taking this into account and expanding the last term in the denominator of (11) in powers of $ \Omega \theta $ we obtain:
 \bea 
 && q v_x (\psi + \Omega \theta)
                   \nn\\
 &\approx& \!\!q v_x (\psi) + \eta \Omega q \tau (1 + iql 
\cos \psi + i k \Omega \tau - i \omega \tau)^{-1} \frac{dv_x}{d \psi}
 \nn\\\nn\\ &
+& \!\! q\frac{\eta^2}{2}(\Omega \tau)^2 (1 + i q l \cos \psi + i k \Omega \tau 
- i \omega \tau)^{-2} \ \frac{d^2 v_x}{d \psi^2}. \,
         \eea
 Substituting this asymptotic expression into (9) we can calculate first terms of the expansions of relevant components of the CF conductivity in powers of the small parameter $ (qR)^{-1} $ where $ R = v_F/\Omega $ is the CF cyclotron radius. Within the "collisionless" limit $ 1/\tau \to 0 $ we have:
 \bea 
 \tilde \sigma_{xx} & =& - N \frac{i\omega}{q^2} e^2 \left \{1 + \frac{i \delta}{\sqrt{1 - \delta^2}} 
+ \frac{ i \delta}{\sqrt{(1 - \delta^2)^5}} \right.
  \nn\\\nn\\
 && \times  \left.
\frac{1}{2(qR)^2} \bigg (1 - \frac{5}{4} \, \frac{1}{1 - \delta^2}
\bigg) \right \};
                 \\\nn\\
 \tilde \sigma_{yy} &=& N \frac{v_F e^2}{q} \left \{\sqrt{1 - \delta^2} + i \delta + \frac{1}{2 (qR)^2}  \right.
     \nn\\\nn\\ && \left.
\times \left [\frac{7}{4} 
\frac{1}{\sqrt{(1 -\delta^2)^5}} - 
\frac{1}{\sqrt{(1 -\delta^2)^3}} \right ] \right \};
                 \\ \nn\\
  \tilde \sigma_{xy} & =& i N \frac{v_F e^2}{q} \frac{\delta}{2 qR} \bigg [\frac{1}{\sqrt{1 - \delta^2}} + \frac{\delta^2}{\sqrt{(1 - \delta^2)^3}}
\bigg ].
      \eea
 Here, $ N = m^*/2\pi \hbar^2 $ is the density of states at the CF Fermi
surface, and $ \delta = \omega /q v_F.$ Using these results we can easily
get approximations for the functions $ K_{\alpha \beta (i)}^0 ({\bf q},
\omega) \ (\alpha, \beta = 0.1)$ and, subsequently, the desired
density-density response function given by (4). It was shown \cite{three}
that the integral over $ \omega $ in the expression for $ \rho_D $ (1) is
dominated by $ \omega \sim T,$ and the major contribution to the integral
over $q$ in this expression comes from $ q \sim k_F (T/T_0 )^{1/3} ,$
where $k_F$ is the Fermi wave vector and the scaling temperature $ T_0 $
is defined below. Therefore we get an estimate for $ \delta, $ namely $
\delta \sim (T/\mu ) (T_0 /T)^{1/3}, $ where $ \mu $ is the chemical
potential of a single 2DEG included in the bilayer. For the parameter  $ T_0
$ taking on values of the order of room temperature, $\delta $ is small compared to unity at low temperatures $
(T \sim 1 K).$

Here,  we limit ourselves to the case of two identical
layers $ (\Pi_{(1)} = \Pi_{(2)} \equiv \Pi) $. 
 For $ \delta \ll 1$ we obtain the approximation: 
   \bea 
 && \Pi_{00} ({\bf q}, \omega ) 
  \nn\\  \nn \\&
= &\frac{q^3}{\displaystyle q^3 \Big (\frac{d n}{d \mu}\Big)^{-1} - 8 \pi i \hbar \omega k_{F} \Big(1 + 2(k_{F} R)^{-1} +
\frac{3}{8} (q R)^{-2} \Big)}.\nn\\
 \eea
 Here, $ dn/d\mu $ is the compressibility of the $ \nu = 1/2 $ state which
is defined as \cite{three}:
 \be 
 \frac{dn}{d \mu} \equiv \Pi_{00} ({\bf q} \to 0; \ \omega \to 0 ) =\frac{3 m^*}{8 \pi \hbar^2 }.
  \ee
 This differs from the compressibility of the noninteracting 2DEG in the
absence of an external magnetic field (the latter equals N). The
difference in the compressibility values is a manifestation of the
Chern-Simons interaction in strong magnetic fields.

 In following calculations we adopt the expression used in the work
\cite{three} for the screened interlayer potential $ U ({\bf q,}\omega),$
namely:
 \bea 
  U ({\bf q,}\omega) &=& \frac{1}{2}\ \frac{V_b + U_b}{1 + \Pi ({\bf q,}\omega) (V_b + U_b)} 
         \nn\\ \nn\\ &&
-\frac{1}{2}\ \frac{V_b - U_b}{1 + \Pi ({\bf 
q,}\omega) (V_b - U_b)}
                 \eea
 where $\ V_b {\bf (q)} = 2 \pi e^2/\epsilon q $
and $\ U_b {\bf (q)} =  (2 \pi e^2/\epsilon q) e^{-q
d} $ are Fourier components of the bare Coulomb potentials for
intralayer and interlayer interactions, respectively, and $ \epsilon $ is
the dielectric constant. Substitung (18) into (1) and using our result
(16) for $ \Pi \bf (q, \omega) $ we can present the transresistvity in the
"ballistic" regime as:
 \be
\rho_D = \rho_{D0} + \delta \rho_D.
   \ee
 Here, the first term $ \rho_{D0} $ is the transresistivity at $ \nu =
1/2 $ when the effective magnetic field is zero, and the second term gives
a correction arising in a nonzero effective magnetic field (away from $
\nu = 1/2 $). As it was to be expected, our expression for $ \rho_{D0} $
coincides with the already known result  \cite{three}: 
        \be 
\rho_{D0} = \frac{h}{e^2} \frac{\Gamma (7/3) \zeta (4/3)}{3 \sqrt 3}
\bigg(\frac{T}{T_0} \bigg)^{4/3}
        \ee
 with $ T_0 = \big(\pi e^2 n d/\epsilon\big) (1 + \alpha),$
and 
  \be
 \displaystyle{\frac{1}{\alpha} = \frac{2 \pi e^2 d}{\epsilon}
\frac{dn}{d \mu}} .
    \ee
 The leading term of the correction $ \delta \rho_D $ at low temperatures  $ (T/T_0)\ll 1 $ can be writen as follows:
 \bea 
 \delta \rho_D &\!
= \!&\! \frac{2}{3} \rho_{D0} \frac{1}{k_F R} \left(1 +
\frac{3}{8} \frac{1}{k_FR}\right)
+ a^2 \frac{h}{e^2} \left(\frac{2T}{T_0} \right)^{2/3}\!\!
\frac{1}{(k_F R)^2}
         \nn\\\nn\\
 &&\!\!\approx \frac{4}{3} \rho_{D0} \Delta \nu \left(1 + \frac{3}{4} \Delta
\nu \right) + 4a^2 \frac{h}{e^2} \left( \frac{2T}{T_0}
\right)^{2/3}\!\! (\Delta \nu)^2 . \nn\\
           \eea  
 Here, the dimensionless positive constant $ a^2 $ can be approximated as:
  \be 
a^2 = \frac{7}{24\sqrt 3} \int_0^\infty \bigg(\frac{y^{2/3}}{\sinh^2 y} - \frac{1}{y^{4/3} \cosh^2 y} \bigg) dy. 
           \ee  

We have to remark that our result (23) cannot be used in the limit $ T \to
0. $ Actually, this expression provides a good asymptotic form for the
coefficient $ a^2 $ when $ (T k_F l / \mu )^{1/3} \geq 1.5 .$ Assuming
that the mean free path is of the order of $ 1.0 \mu m $ as in the
experiments  \cite{eleven} on dc magnetotransport in a single modulated
2DEG at $ \nu $ close to $1/2,$ and using the estimate of \cite{seven} for
the electron density $ n = 1.4 \times 10^{15} m^{-2} ,$ we obtain that the
expression (23) gives good approximation for $ a^2 $ when $ T/\mu $ is no less than $10^{-2}$.

It follows from our results (19), (22) that transresistivity $
\rho_D $ enhances nearly quadratically with $ \Delta \nu $ when the
filling factor deviates from $ \nu = 1/2 .$ The linear in $\Delta \nu $
term is also present in the expression for $ \delta \rho_D. $ This causes
an asymmetric shape of the plot of Eq. (22) relative to $ \Delta \nu = 0.
$ However, this asymmetry is not very significant for the linear term is
smaller than the last term on the right hand side of (22). This difference
in magnitudes is due to different temperature dependences of the
considered terms. The first term  including the linear in $ (k_F
R)^{-1}$ correction is proportional to $ (T/T_0)^{4/3},$ whereas the
second one is proportional to $ (T/T_0)^{2/3}$ and predominates at low
temperatures. 
 So, the magnetic field dependence of the transresistivity near $\nu = 1/2 
$ matchs that observed in the experiments (See Fig. 1).

\begin{figure}[t]
\begin{center}
\includegraphics[width=7.8cm,height=8.7cm]{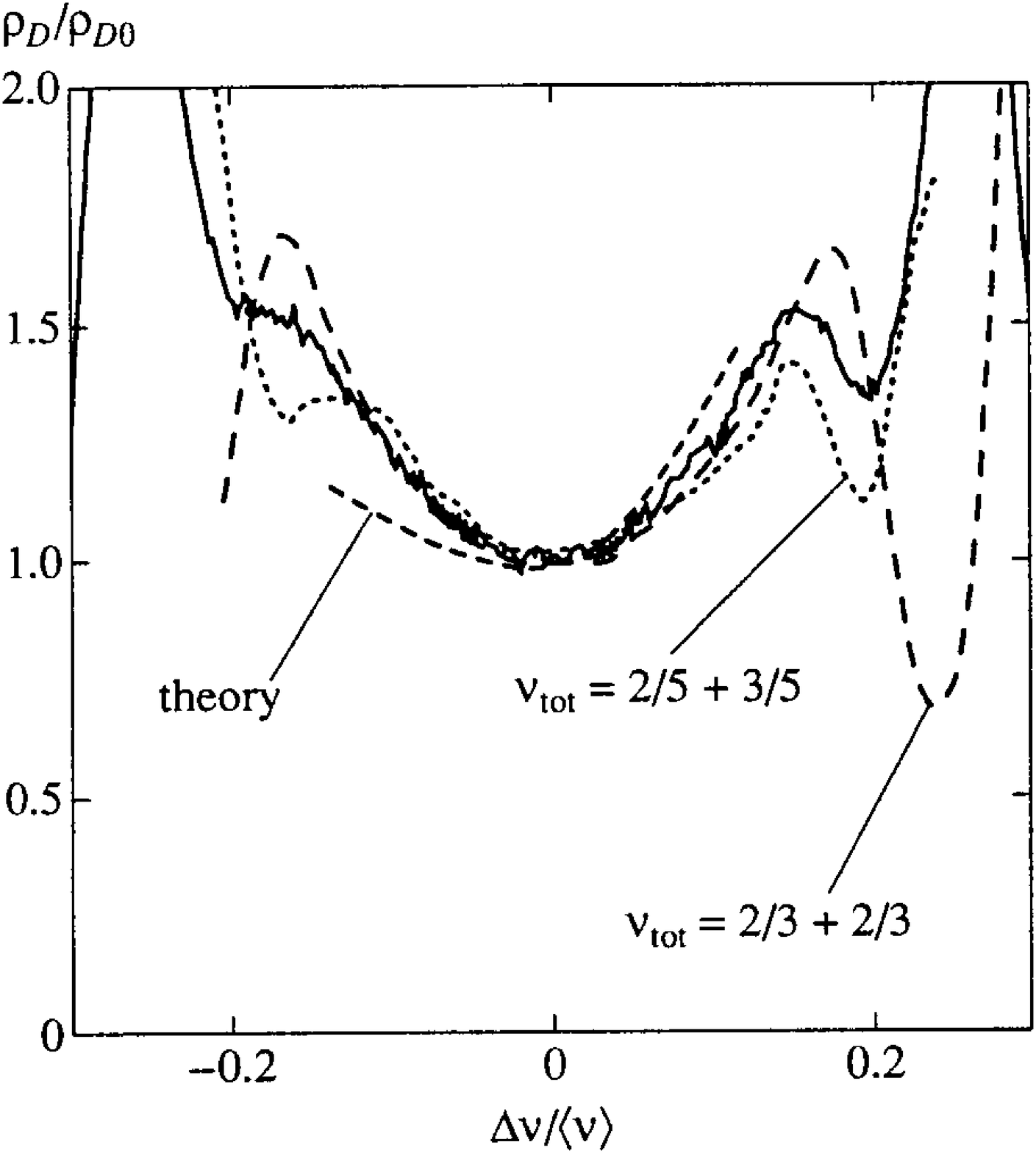}
\caption{
Scaled drag resistivity versus $ \Delta \nu $ at T = 0.6; lowest dashed curve is the plot of Eq. (22) at $ m^* = 4 m_b; \ A_0 = 15, $ and remaining curves present experimental data of [7];
}  
\label{rateI}
\end{center}
\end{figure}

Keeping only the greatest term in (22), the ratio $ \rho_D /\rho_{D0}$ can
be presented in the form:
           \be  
\frac{\rho_D}{\rho_{D0}} = 4 \beta (\Delta \nu)^2 + 1.
            \ee  
 Here, the coefficient $ \beta $ equals:
           \be
 \beta = \frac{3 \sqrt 3 a^2} {\Gamma(7/3) \zeta(4/3)}
\left(\frac{2T_0}{T} \right)^{2/3}.
           \ee  
 This coefficient is proportional to the curvature of the plot of Eq. (22)
assuming that the first term is neglected. The curvature reveales a
strong dependence on temperature whose character also agrees with
experiments of \cite{seven} as it is shown in Fig. 2.

\begin{figure}[t]
\begin{center}
\includegraphics[width=8cm,height=8.4cm]{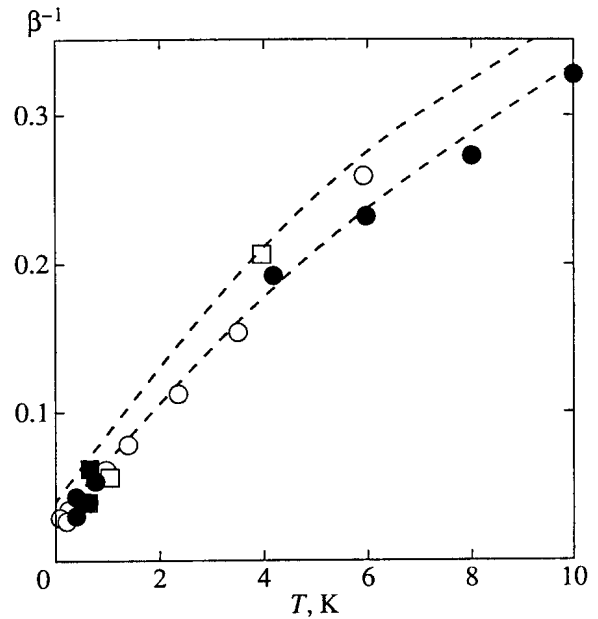}
\caption{Temperature dependence of the coefficient $ \beta^{-1} $ for
interlayer distances $ d = 10 nm $ (upper curve) and $ d = 22.5 nm $
(lower curve) compared to the summary of experimental curvature at both
spacings [7]}
\label{rateI}
\end{center}
\end{figure}

A striking feature in the experimental results is that they appear to be
nonsensitive to the distance between the 2DEGs. Sets of data corresponding
to samples with different interlayer spacings $ d_A = 10 nm$ and $ d_B =
22.5 nm$ fall on the same curve. This concerns both magnetic field
dependence of the transresistivity and temperature dependence of the
parameter $\beta$.  Results of the present analysis provide a possible
explanation for this feature. It follows from (20)--(25) that the
dependence of $ \rho_D$ of the interlayer spacing is completely included
in the characteristic temperature $ T_0 $ which is defined with Eq. (21).
The above quantity is nearly independent of the interlayer separation $ d
$ when the parameter $ \alpha $ takes on values larger that unity.
 Estimating the parameter $ \alpha $ as it is
given by Eq. (21), we obtain that the condition $ \alpha > 1 $ could be
satisfied for small values of the  compressibility of the $ \nu = 1/2 $ 
state. 
  However, within the RPA the effective mass of CFs coincides with the
single
electron band mass $ m_b $ which takes on the value $ m_b \approx 0,07 m_e
$ for GaAs wells ($ m_e $ is the mass of a free electron). Using this
value to estimate the compressibility  as it is introduced by Eq. (17)
we get $ \alpha \approx 0.44. $ This is too small to provide insensitivity
of the coefficient $ \beta $ determined by Eq. (25) to the interlayer
distance for interlayer spacings reported  in the experiments
\cite{three}. The above discrepancy
could be removed taking into account Fermi liquid interactions among
quasiparticles (CFs).  To include Fermi liquid effects into consideration
we write the renormalized polarizability $ \Pi^* $ in the form
\cite{eight}:
 \be
\Pi^{* -1} = \Pi^{-1} + F_{(0)} + F_{(1)}.
 \ee  
 Here, $\Pi $ is the polarizability of noninteracting CFs defined with Eq.
(2), and the remaining terms present contributions arising due to Fermi
liquid interaction in the CF system. Only contributions from the first and
greatest two terms in the expansion of the Fermi liquid interaction
function in Legendre polynomials ($f_0$ and $ f_1$, respectively) are kept
in Eq. (26) to avoid too lengthy calculations. Matrix elements of the $
2\times 2 $ matrices $ F_{(0)}$ and $ F_{(1)}$ equal:
\bea 
 F_{(0)} &= &
\left( \begin{array}{cc}
f_0 &   0 \nn \\
 0 &  0
\end{array}
\right) 
  \\   \\   
F_{(1)} & = &
\left( \begin{array}{cc}
{\ds \frac{m^* - m_b}{ne^2}  \frac{\omega^2}{q^2} } & 
 0 \nn\\
0 &  {-\ds \frac{m^* - m_b}{ne^2}}\\
\end{array} \right) .
 \eea

 Within the Fermi liquid theory the effective mass $ m^* $ is related to
the "bare" mass $ m_b $ as follows:
 \be
 \frac{1}{m_b} = \frac{1}{m^*} + \frac{f_1}{2 \pi \hbar^2} \equiv
\frac{1 + A_1 }{m^*}.
 \ee
 Using these expressions (26)--(28) and carrying out calculations within
the relevant limit $ \delta \ll 1 ,$ we obtain that the expression for the
density-density response function for a single layer keeps the form given
by Eq. (16) where the compressibility $ dn/d\mu $ is replaced with the
quantity $ dn^*/d\mu $ renormalized due to the Fermi liquid interaction:
 \be
\frac{dn^*}{d \mu} 
  = \frac{3m^*}{8 \pi \hbar^2} \bigg (1 + \frac{3m^*}{8 \pi \hbar^2} 
f_0 \bigg )^{-1}  \!\!\equiv \frac{dn}{d\mu} \bigg( 1+ \frac{dn}{d\mu} f_0
\bigg )^{-1} \!.
   \ee

For strongly correlated quasiparticles this renormalization may
significantly reduce the compressibility of the CF liquid, and,
consequently, increase the value of the parameter $ \alpha. $ It is
usually assumed \cite{three,eight} that the Fermi liquid renormalization
of the effective mass significantly changes its value: $ m^* \sim 5-10 \
m_b. $ This gives for the Fermi liguid coefficient $A_1$ values of the
order of 10.  Using this estimate, and substituting our renormalized
compressibility (29) into the expression (21) we arrive at the conclusion
that $dn^*/d\mu $ is low enough for the condition $ \alpha > 1, $ to be
satisfied when the Fermi liquid parameter $ A_0 \equiv f_0 / 2\pi \hbar^2
$ takes on values of the order of $10-100.$ This conclusion does not seem
an unrealistic one for it is reasonable to expect $ A_0 $ to be of the
order or greater than the next Fermi liquid parameter $ A_1.$
 We obtain
a reasonably good agreement between the plot of our Eq. (22) and the
experimental results, using 
  $A_0 = 15$ and $A_1 = 3 \ (m^* = 4 m_b).$ (Fig. 1).

 Our results for temperature dependence of $ \beta^{-1}$ also agree with
the results of experiments \cite{seven}. The upper curve in Fig.2
corresponds to the double-layer system with with smaller interlayer
spacing $ d_A = 10 nm $ which gives $ T_0 = 487 K,$ and the lower curve
exhibits the temperature dependence of $ \beta^{-1} $ for greater spacing
$ d_B = 22.5 nm \ (T_0 = 587 K).$ The curves do not coincide but they are
arranged rather close to each other.

Finally, the results of the present analysis enable us to qualitatively
describe all important features observed in experiments of \cite{seven} on
the Coulomb drag slightly away from one half filling of lowest Landau
levels of both interacting 2DEG. 
   They also give us grounds to treat these experimental results as one
more evidence of strong Fermi liquid interaction in the CF system near one
half filling of the lowest Landau level. The above interaction provides a
significant reduction of the compressibility of the CF liquid and a
consequent enhancement in the screening length in single layers. 
Essentially, the parameter $ \alpha $ characterizes the ratio of the
Thomas--Fermi screening length in a single 2DEG at $ \nu = 1/2 $ and the
separation between the layers \cite{three}. When $ \alpha > 1, $
intralayer interactions predominate those between the layers which could
be the reason for low sensitivity of the bilayer to changes in the
interlayer spacing. It is likely that here is an explanation for the
reported nearly independence of the drag on the interlayer separation
\cite{seven}. We believe that at larger distances between the layers the
dependence of the transresistivity of $ d $ could be revealed in the
experiments. At the same time the results of \cite{seven} give us a
valuable opportunity to estimate a strength of Fermi liquid interactions
between quasiparticles at $ \nu = 1/2 $ state which is important for
further studies of such systems.

\vspace{0mm}

{\it  Acknowledgments:}
The author thank K.L. Haglin and G.M. Zimbovsky for help with the manuscript.



\end{document}